\documentclass{aa}  
\usepackage{graphicx}

\usepackage{txfonts}

\def\zl{z$_{\rm lens}~$}

\begin{document}

   \title{COSMOGRAIL: the COSmological MOnitoring of \\
    \vspace*{1mm}  GRAvItational  Lenses   \thanks{Based    on
    observations made with the ESO-VLT  Unit Telescope 2 Kueyen (Cerro
    Paranal, Chile; Programs 077.A-0155, PI: G. Meylan).}
    }

   \subtitle{VI. Redshift of the lensing galaxy in 
    seven gravitationally lensed quasars}

   \titlerunning{COSMOGRAIL: VI. Redshift of the lensing galaxy in 
   seven gravitationally lensed quasars}

   \author{A. Eigenbrod\inst{1} \and F. Courbin\inst{1} 
        \and G. Meylan\inst{1} 
	}
    
    %
    %
   

    \institute{
     Laboratoire d'Astrophysique, Ecole Polytechnique F\'ed\'erale
     de Lausanne (EPFL), Observatoire, 1290 Sauverny, Switzerland
     }

   \date{Received ... ; accepted ...}

 
  \abstract
   {}
   {The knowledge of the redshift of a lensing galaxy that produces multiple images of a
background quasar is essential to any subsequent modeling, whether related 
to the determination of the Hubble constant H$_0$  or to the mass 
profile of the lensing galaxy. 
   We  present the results of our ongoing spectroscopic  observations of gravitationally
lensed quasars in order to measure the  redshift of  their  lensing
galaxies.  We report on the determination  of the lens redshift in  seven
gravitationally lensed systems.   }
   {Our deep VLT/FORS1 spectra are spatially deconvolved
   in order to separate the spectrum of the  lensing galaxies from the
   glare of the much brighter quasar images. Our observing  strategy
   involves   observations in Multi-Object-Spectroscopy (MOS) mode    which
   allows the simultaneous  observation  of the target and  of several crucial
   PSF  and flux calibration stars. The  advantage of this method over
   traditional  long-slit observations is that it allows a  much  more reliable
   extraction and flux calibration of the spectra.}
   {We obtain the  first reliable  spectra of the lensing 
    galaxies in six 
    lensed quasars:  
    FBQ~0951$+$2635   (\zl$=0.260$), 
    BRI~0952$-$0115   (\zl$=0.632$),
    HE~2149$-$2745    (\zl$=0.603$), 
    Q~0142$-$100      (\zl$=0.491$),
    SDSS~J0246$-$0825 (\zl$=0.723$), and
    SDSS~J0806$+$2006 (\zl$=0.573$). The  last three redshifts also correspond
    to the \ion{Mg}{II} doublet seen in absorption in the quasar spectra 
    at the lens redshift.  
    Our spectroscopic redshifts of HE~2149$-$2745 and FBQ~0951$+$2635 are 
    higher than previously reported, which means that H$_0$ estimates from 
    these two systems must be revised to higher values.
    Finally, we reanalyse the blue side of our previously published 
    spectra of Q~1355$-$2257 and find \ion{Mg}{II} in absorption
    at z~$=0.702$, confirming our previous redshift estimate.
    The spectra of all lenses are typical of early-type galaxies.} 
   {}


   \maketitle

\section{Introduction}

About 100  gravitationally lensed quasars  have  been found  since the
discovery of  the first case by Walsh   et al.  (\cite{walsh79}).  An
advantage of lensed quasars resides in their possible variability, 
potentially leading to the measurement of the so-called time
delay between    the lensed images of  the   source.  This quantity is
directly related to the  Hubble constant  H$_0$   and to the slope of
the mass profile  of the lensing galaxy  at the position of the quasar
images projected on the plane of the sky. Assuming a model for the mass
distribution in the lensing galaxy,  and measuring the time delay, one
can infer the  value of H$_0$ (Refsdal \cite{Refsdal64}). Conversely,
for a given H$_0$ estimate, the mass distribution in the lens can be
reconstructed   from the time  delay  measurement.  In both cases, one
wishes to construct   a large,  statistically  significant sample   of
lensed quasars,  either to reduce the   random errors on  H$_0$, or to
build a large sample of massive early-type galaxies, the mass profile
of which can be strongly constrained, thanks to gravitational lensing.

For whatever subsequent application of quasar lensing, some  of the key
measurements to carry out in each individual system  are (i) the value of
the time delay, (ii) the astrometry of all quasar images with respect to the 
lensing galaxy, and (iii) the redshift of the lensing galaxy. 

The     main    goal     of    COSMOGRAIL   (e.g.       Eigenbrod   et
al.~\cite{cosmograil1})  is to measure  a large  number of time delays
from  a photometric  monitoring campaign based on a few 2-m class telescopes.  
While this goal
can be reached only in the long run, it is possible to pave the way to
accurate  modeling of the  systems  by obtaining  the redshift of  the
lensing  galaxies in all the systems  currently monitored. This is the
goal   of the   present  paper, which  is  the   continuation  of  our
spectroscopic  study of lensed   quasars, undertaken  at the  VLT with
FORS1 (e.g. Eigenbrod et al.~\cite{cosmograil3}).

\section{VLT Spectroscopy}

\subsection{Observations}

We present new observations  of six gravitationally lensed quasars, in
order  to  determine   the redshift of   the   lensing galaxy, plus  a
re-analysis  of  one object  previously  published  in  Eigenbrod   et
al.~(\cite{cosmograil3}). The details of the observational setup and of
the   data  reduction  can    be     found   in  Eigenbrod   et    al.
(\cite{cosmograil2}  \& \cite{cosmograil3},),  but  we  remind in  the
following some of the most important points.

Our   observations  are  acquired  with  the   FOcal  Reducer and  low
dispersion  Spectrograph   (FORS1), mounted  on   the ESO  Very  Large
Telescope at the Observatory of Paranal (Chile). 
All  the observations  are carried  out in   the MOS mode
(Multi Object Spectroscopy).  This  strategy is the most convenient to
get  simultaneous observations of the  main target and of several stars
used both as flux calibrators and as reference point-spread functions
(PSF) in order to spatially deconvolve the spectra.
We choose these stars to be located as close as possible to 
the gravitationally lensed quasars with similar apparent magnitudes.

All targets are observed with the high-resolution collimator, allowing
us to observe  simultaneously eight  objects over  a field  of view of
$3.4\arcmin\times3.4\arcmin$ with a  pixel scale of $0.1\arcsec$.  The
GG435 order sorting filter in combination with the G300V grism is used
for   all    objects,       giving   a   useful      wavelength  range
4450$~<\lambda<~$8650~\AA\ and a scale of $2.69$~\AA\ per pixel in the
spectral  direction.  This     setup   has a   spectral     resolution
$R=\lambda/\Delta \lambda  \simeq  200$  at   the  central  wavelength
$\lambda=5900$~\AA\ for  a $1.0\arcsec$ slit  width in the case of the
high resolution collimator.  The  choice of this grism favors spectral
coverage rather than spectral resolution as we aim at measuring
unknown lens redshifts.

\begin{table}[t!]
\caption[]{Journal of the observations}
\label{refer}
\begin{flushleft}
\begin{tabular}{cccccc}
\hline 
\hline 
ID & Date & Seeing $[\arcsec]$ & Airmass & Weather \\
\hline 
\multicolumn{3}{l}{Q~0142$-$100} \\
\hline 
1 & 11/08/2006 & 0.83 & 1.119 &   Photometric\\
2 & 11/08/2006 & 0.77 & 1.078 &   Photometric\\
3 & 19/08/2006 & 0.78 & 1.709 &   Photometric\\
4 & 19/08/2006 & 0.79 & 1.507 &   Photometric\\
\\ 
\hline 
\multicolumn{3}{l}{SDSS~J0246$-$0825} \\
\hline 
1 & 22/08/2006 & 0.76 & 1.490 &   Photometric\\
2 & 22/08/2006 & 0.67 & 1.350 &   Photometric\\
3 & 22/08/2006 & 0.60 & 1.234 &   Photometric\\
4 & 22/08/2006 & 0.58 & 1.161 &   Photometric\\
5 & 22/08/2006 & 0.61 & 1.103 &   Photometric\\
6 & 22/08/2006 & 0.59 & 1.069 &   Photometric\\
\\ 
\hline 
\multicolumn{3}{l}{SDSS~J0806$+$2006} \\
\hline 
1 & 22/04/2006 & 0.86 & 1.442 &   Photometric\\
2 & 22/04/2006 & 0.90 & 1.494 &   Photometric\\
3 & 23/04/2006 & 0.93 & 1.590 &   Photometric\\
4 & 23/04/2006 & 0.95 & 1.714 &   Photometric\\
\\ 
\hline 
\multicolumn{3}{l}{FBQ~0951$+$2635} \\
\hline 
1 & 31/03/2006 & 0.68 & 1.600 &   Photometric\\
2 & 31/03/2006 & 0.74 & 1.593 &   Photometric\\
3 & 01/04/2006 & 0.59 & 1.598 &   Photometric\\
4 & 01/04/2006 & 0.57 & 1.629 &   Photometric\\
\\ 
\hline 
\multicolumn{3}{l}{BRI~0952$-$0115} \\
\hline 
1 & 23/04/2006 & 0.67 & 1.097 &   Photometric\\
2 & 24/04/2006 & 0.58 & 1.087 &   Photometric\\
3 & 24/04/2006 & 0.56 & 1.094 &   Photometric\\
4 & 24/04/2006 & 0.50 & 1.116 &   Photometric\\
5 & 24/04/2006 & 0.42 & 1.165 &   Photometric\\
6 & 24/04/2006 & 0.45 & 1.225 &   Photometric\\
\\ 
\hline 
\multicolumn{3}{l}{Q~1355$-$2257} \\
\hline 
1 & 05/03/2005 & 0.68 & 1.016 & Photometric\\
2 & 05/03/2005 & 0.73 & 1.040 & Photometric\\
3 & 20/03/2005 & 0.63 & 1.038 & Photometric\\
4 & 20/03/2005 & 0.54 & 1.015 & Photometric\\
5 & 20/03/2005 & 0.57 & 1.105 & Photometric\\
6 & 20/03/2005 & 0.56 & 1.166 & Photometric\\
\\ 
\hline 
\multicolumn{3}{l}{HE~2149$-$2745} \\
\hline 
1 & 04/08/2006 & 0.66 & 2.004 &    Photometric\\
2 & 04/08/2006 & 0.62 & 1.724 &    Photometric\\
3 & 04/08/2006 & 0.62 & 1.461 &    Photometric\\
4 & 04/08/2006 & 0.69 & 1.328 &    Photometric\\
5 & 04/08/2006 & 0.52 & 1.204 &    Photometric\\
6 & 04/08/2006 & 0.59 & 1.134 &    Photometric\\
\hline 
\end{tabular}
\end{flushleft}
\end{table}
 
\begin{figure}[t!]
\begin{center}
\includegraphics[width=8cm]{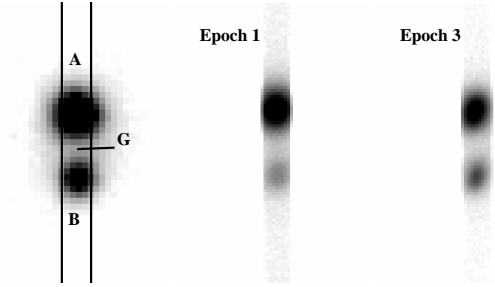}
\caption{Q~0142$-$100.  Slit width: 1.0\arcsec. Mask PA : $+75^{\circ}$}
\label{Q0142_slits}
\end{center}
\end{figure}

\begin{figure}[t!]
\begin{center}
\includegraphics[width=8cm]{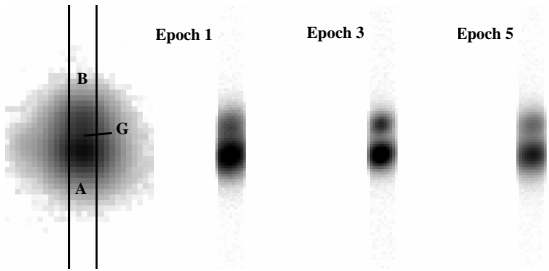}
\caption{SDSS~J0246$-$0825.  Slit width: 1.0\arcsec. Mask PA : $+55^{\circ}$}
\label{SDSS0246_slits}
\end{center}
\end{figure}

\begin{figure}[t!]
\begin{center}
\includegraphics[width=8cm]{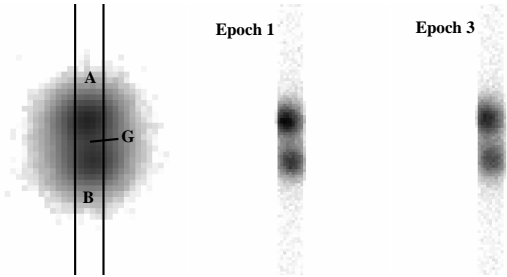}
\caption{SDSS~J0806$+$2006.  Slit width: 1.0\arcsec. Mask PA : $-56^{\circ}$}
\label{SDSS0806_slits}
\end{center}
\end{figure}

\begin{figure}[t!]
\begin{center}
\includegraphics[width=8cm]{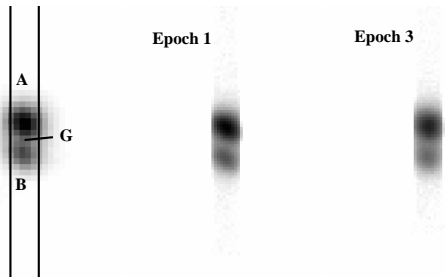}
\caption{FBQ~0951$+$2635.  Slit width: 1.0\arcsec. Mask PA : $+53^{\circ}$}
\label{FBQ0951_slits}
\end{center}
\end{figure}

\begin{figure}[t!]
\begin{center}
\includegraphics[width=8cm]{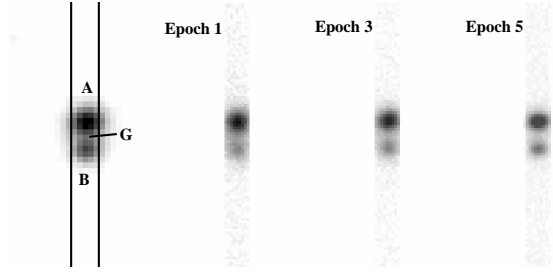}
\caption{BRI~0952$-$0115.  Slit width: 1.0\arcsec. Mask PA : $-45^{\circ}$}
\label{BRI0952_slits}
\end{center}
\end{figure}

\begin{figure}[t!]
\begin{center}
\includegraphics[width=8cm]{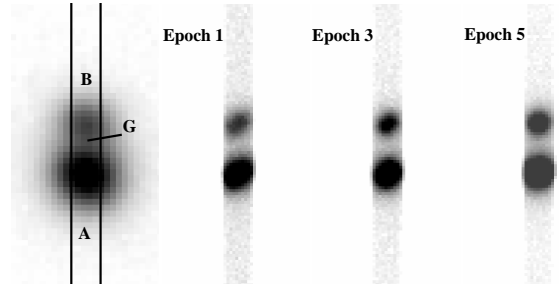}
\caption{HE~2149$-$2745.  Slit width: 1.0\arcsec. Mask PA : $-32^{\circ}$}
\label{HE2149_slits}
\end{center}
\end{figure}

We    choose  slitlets  of $1.0\arcsec$  width,    matching the seeing
requested for these service-mode observations. Our observing sequences
consist of a short acquisition image, an ``image-through-slit'' check,
followed  by   two consecutive  deep   spectroscopic  exposures.   All
individual exposures for all  objects are $1400$~s long.  The journals
of the observations are given in Table~\ref{refer}.   
The   through-slit  images   are  displayed  in
Figs.~\ref{Q0142_slits} to  \ref{HE2149_slits}, where the epochs refer
to     the   exposure     numbers   in  Table~\ref{refer}.

For every object we center at least two   slitlets on foreground stars
and one slitlet along  the lensed images of the  quasar.  The mask  is
rotated to a Position Angle that  avoids clipping of any quasar image.
This  is mandatory to carry out  spatial deconvolution of the spectra.
The spectra of the PSF stars are also used to cross-calibrate the flux
scale as  the data are taken at  different airmasses  and at different
epochs      (see      Eigenbrod  et     al.~\cite{cosmograil2}).   Six
spectrophotometric  standard stars are  used to carry out the relative
flux calibration, i.e.  GD~108, HD~49798, LTT~377, LTT~1020, LTT~1788,
and LTT~7987.

\subsection{Reduction and Deconvolution}

We  follow exactly the  same  procedure as described  by  Eigenbrod et
al. (\cite{cosmograil3}).  We carry out  the standard bias subtraction
and flat  field correction of the  spectra using IRAF\footnote{IRAF is
distributed by the National Optical Astronomy Observatories, which are
operated by the Association of Universities for Research in Astronomy,
Inc.,    under cooperative    agreement  with  the    National Science
Foundation.}.
The wavelength calibration is obtained from the
spectrum of helium-argon lamps.  
The sky    background  and cosmic  rays   are  removed using  multiple
exposures.

A flux  cross-calibration of the spectra  is applied  before combining
them  into  one final spectrum.   This  is done efficiently using  the
spectra of the PSF stars as references.

\begin{table}[t!]
\caption[]{Redshift values determined for the lensing galaxies in the seven
gravitational lenses.}
\label{redshift}
\begin{flushleft}
\begin{tabular}{lc}
\hline 
\hline 
Object                 & z$_{\rm lens}$   \\
\hline 
Q~0142$-$100	       & $0.491 \pm 0.001$\\
SDSS~J0246$-$0825      & $0.723 \pm 0.002$\\
SDSS~J0806$+$2006      & $0.573 \pm 0.001$\\
FBQ~0951$+$2635        & $0.260 \pm 0.002$\\  
BRI~0952$-$0115        & $0.632 \pm 0.002$\\
Q~1355$-$2257          & $0.702 \pm 0.001$\\
HE~2149$-$2745         & $0.603 \pm 0.001$\\
\hline  	     
\end{tabular}	     
\end{flushleft}
\end{table}

In order to separate  the spectrum of the  lensing galaxy from  
the spectra of the much brighter quasar  images, we use the  spectral  
version of the MCS deconvolution  algorithm 
(Magain, Courbin \& Sohy \cite{magain98},  Courbin et al.  \cite{courbin})
  
This software    uses  the spatial   information
contained  in the spectra of several reference PSF stars.  
The deconvolved  spectra are  sharpened in the spatial direction,
and also  decomposed   into a ``point-source  channel'' containing  the
spectra of the  quasar images, and an ``extended  channel'' containing
the spectra of everything  in the image  which is not a point-source, 
i.e. in this case the spectrum of the lensing galaxy.

\begin{figure}[t!]
\begin{center}
\includegraphics[width=8.7cm]{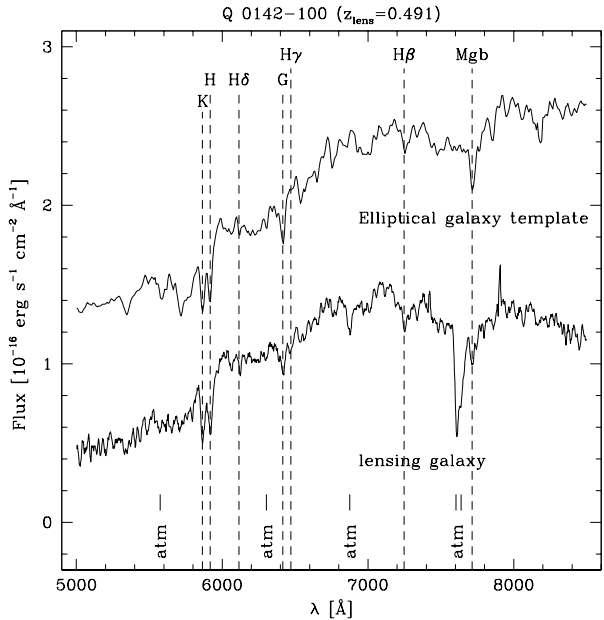}
\caption{Spectrum of the lens in Q~0142$-$100. The
total integration time is 5600~s. The template spectrum of a redshifted 
elliptical galaxy  is shown for comparison (Kinney et al. \cite{kinney}).
Atmospheric absorptions are indicated in all figures
by the  label {\tt atm}.
}
\label{Q0142_lens}
\end{center}
\end{figure}

\begin{figure}[t!]
\begin{center}
\includegraphics[width=8.7cm]{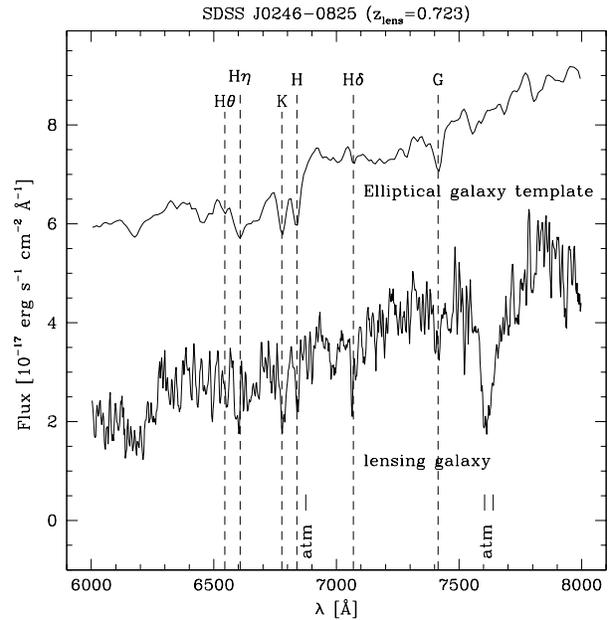}
\caption{Spectrum of the lens in SDSS~J0246$-$0825. The
total integration time is 8400~s.}
\label{SDSS0246_lens}
\end{center}
\end{figure}

\begin{figure}[t!]
\begin{center}
\includegraphics[width=8.7cm]{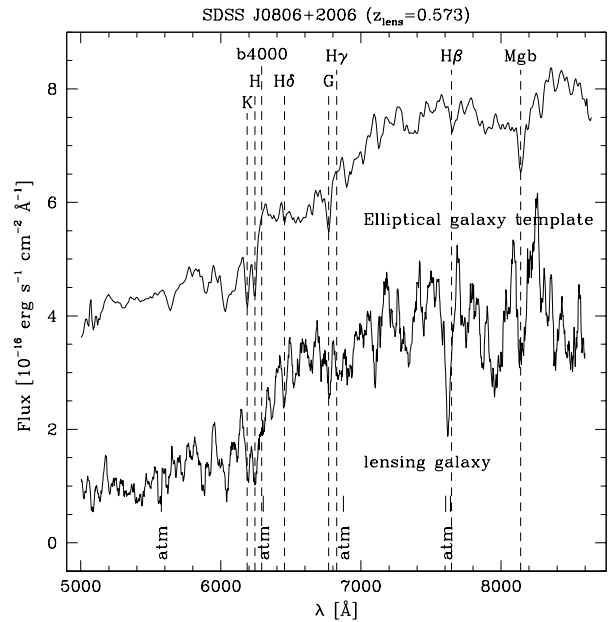}
\caption{Spectrum of the lens in SDSS~J0806$+$2006. The
total integration time is 5600~s.}
\label{SDSS0806_lens}
\end{center}
\end{figure}

\begin{figure}[t!]
\begin{center}
\includegraphics[width=8.7cm]{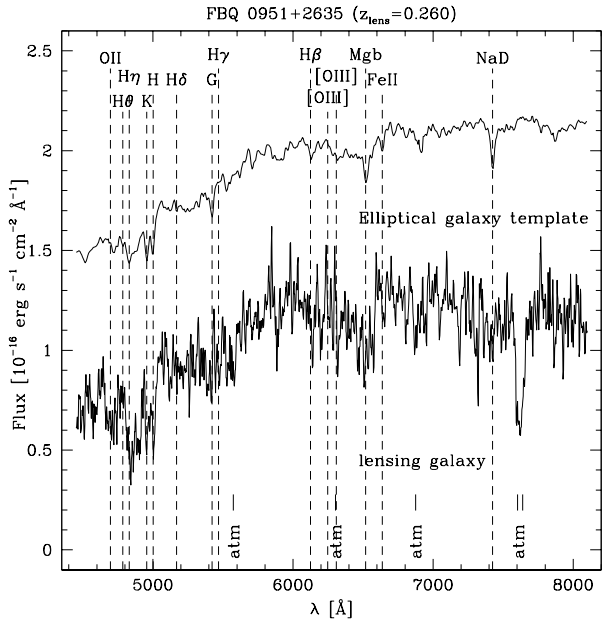}
\caption{Spectrum of the lens in FBQ~0951$+$2635. The
total integration time is 5600~s.}
\label{FBQ0951_lens}
\end{center}
\end{figure}

\begin{figure}[t!]
\begin{center}
\includegraphics[width=8.7cm]{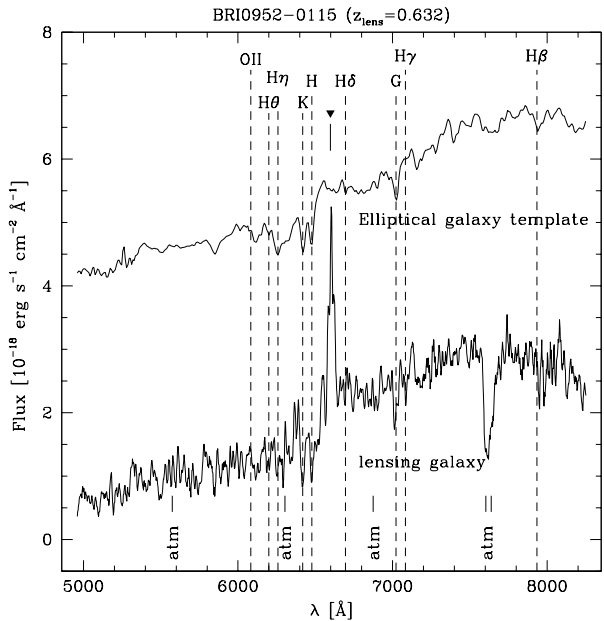}
\caption{Spectrum of the lens in BRI~0952$-$0115. The
total integration time is 8400~s. The emission feature marked by 
the black triangle is residual light of the quasar images.
The Ly$\alpha$ emission of the quasar falls exactly at this wavelength.}
\label{BRI0952_lens}
\end{center}
\end{figure}

\begin{figure}[t!]
\begin{center}
\includegraphics[width=8.7cm]{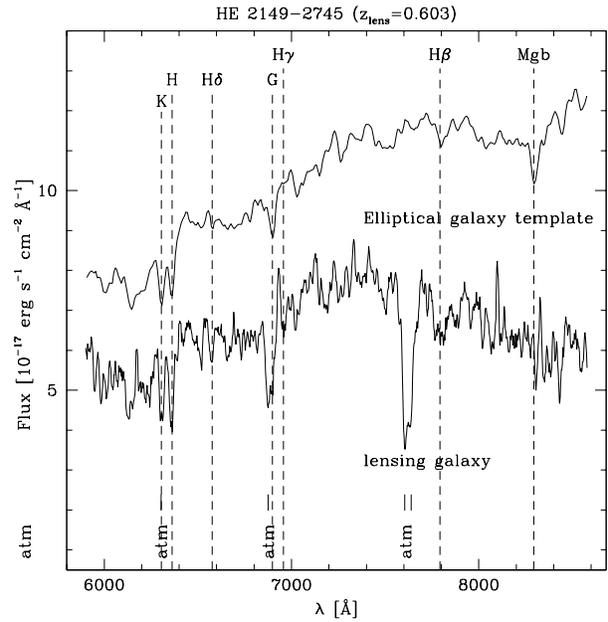}
\caption{Spectrum of the lens in HE~2149$-$2745. The
total integration time is 8400~s.}
\label{HE2149_lens}
\end{center}
\end{figure}

The deconvolved   spectra of the lensing   galaxies are extracted and
smoothed    with   a   10-\AA\   box.     Fig.~\ref{Q0142_lens}  to
\ref{HE2149_lens} display  the  final one-dimensional  spectra, where
the \ion{Ca}{II} H \& K absorption lines  are obvious, as well as the 
$4000$-\AA $\,$  Balmer break, and the G band typical for  CH absorption. 
In some cases, we identify a few more features that are labeled in the
individual figures. The identified lines   are used to  determine the
redshift of the lensing galaxies   given in Table~\ref{redshift}.   We
compute the 1-$\sigma$ error as the standard deviation between all the
measurements of  the individual lines.
The  absence of  emission  lines  in all spectra indicates that  these  
observed  lensing galaxies  are  gas-poor early-type galaxies. 

In most cases, no trace of the quasar  broad emission lines is seen in
the  spectrum of  the  lensing  galaxy,   indicative   of  an accurate
decomposition of the  data into the extended  (lens)  and point source
(quasar images) channels.  Only our VLT spectrum of the lensing galaxy
of BRI~0952$-$0115  is  suffering from residuals  of  the  quasar 
Ly$\alpha$ emission: the presence of the strong Ly$\alpha$ 
in the blue side of the spectrum complicates the deconvolution process.

\begin{figure}[t!]
\begin{center}
\includegraphics[width=8.7cm]{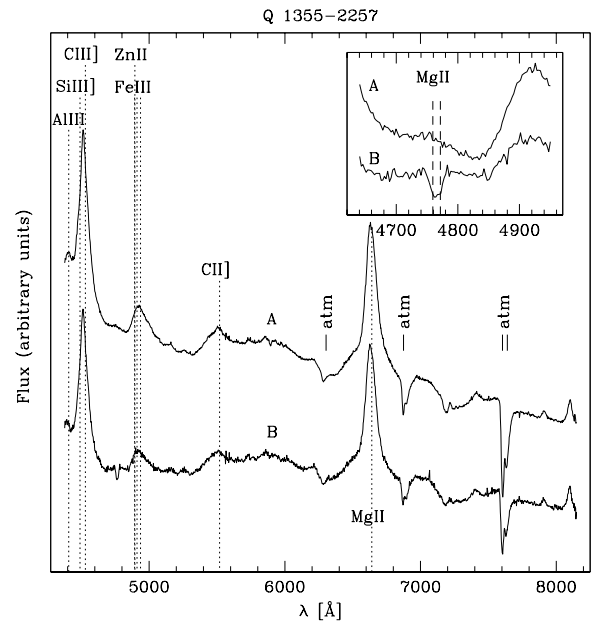}
\caption{Spectrum of the images of quasar Q~1355$-$2257. 
The total integration time is 8400~s.
In the inlet, we show the \ion{Mg}{II}
absorption lines present in the spectrum of component B and which is due to the lensing galaxy.}
\label{Q1355_lens}
\end{center}
\end{figure}

\section{Notes on individual objects}

{\bf Q~0142$-$100  (UM~673  or  PHL~3703)}:  This object   was   first
discovered by Macalpine    \&  Feldman (\cite{macalpine}) as  a   high
redshift quasar  with  z~$=2.719$.   A  few  years  later  Surdej   et
al. (\cite{surdej87})  identified Q~0142$-$100 as a gravitational lens
with two quasar components separated by  $2.2$\arcsec.  No spectrum of
the lensing galaxy has ever been obtained, but
\ion{Ca}{II} and \ion{Na}{I} absorption lines are 
detected in  the spectrum  of the  fainter  quasar image B, suggesting
that  it is    located at a  redshift  \zl$=   0.493$ (Surdej  et  al.
\cite{surdej88}). From our  spectrum of the 
lensing galaxy we confirm that it is the object responsible for the 
quasar absorption lines. Our best estimate of the redshift is 
\zl$= 0.491 \pm 0.001$. The observed galaxy spectrum is typical for
an early type galaxy.

{\bf SDSS~J0246$-$0825}: This double-image  quasar was discovered  in
the course of the  Sloan Digital  Sky  Survey (SDSS)  by Inada et  al.
(\cite{inada05}). The  quasar has  a redshift z~$=1.68$ and an
image   separation of    $1.04$\arcsec.  From   the \ion{Mg}{II}   and
\ion{Mg}{I}  absorption  lines observed in the  spectra  of the two quasar
components, Inada  et al.  (\cite{inada05})  estimate the  redshift of
the lensing galaxy to be
\zl$=0.724$. Our direct redshift measurement \zl$=0.723 \pm 0.002$ 
is in very good agreement  with this value. The spectrum matches well 
that of an elliptical galaxy.

{\bf SDSS~J0806$+$2006}: This  two-image gravitationally lensed quasar
(z~$=1.540$) was  recently discovered by Inada et
al. (\cite{inada06}).  The two  quasar images are separated by
$1.40$\arcsec.  Several  absorption
lines are found in the quasar spectra, such as the (\ion{Mg}{II},
\ion{Ca}{II}~H\&K and \ion{Fe}{II}) lines, at a redshift of \zl$=0.573$. 
The redshift we obtain from our  spectrum is
\zl$=0.573\pm 0.001$. The spectrum 
(see Fig.~\ref{SDSS0806_lens}) matches very well the template spectrum
of an elliptical galaxy.

{\bf FBQ~0951$+$2635  }: A double-image  quasar  at z~$=1.246$  with
$1.1$\arcsec\ image separation found in the course of the FIRST Bright
QSO  Survey (FBQS) by Schechter et   al. (\cite{schechter98}).  A time
delay estimate of $\Delta t=16\pm2$ days  between the quasar images is
given by Jakobsson et  al.  (\cite{jakobsson05}), but the redshift  of
the lensing  galaxy, by the   same authors, is   found to be  elusive.
Schechter  et   al.   (\cite{schechter98})  detect    absorption lines
attributed to \ion{Mg}{II} at z~$=0.73$ and z~$=0.89$ in both spectra
of the quasar  components.  From  the position   of  the lens on   the
fundamental plane,  Kochanek  et  al.   (\cite{kochanek00})  suggest
\zl$\simeq0.21$. We   measure
\zl$=0.260 \pm 0.002$  for this early-type lensing galaxy.
This also corresponds to the photometric redshift of 0.27 measured 
by Williams et al. (\cite{williams}) 
for a group of galaxies in the field of the lensing galaxy.

{\bf  BRI~0952$-$0115}:  This gravitational  lens   was  discovered by
McMahon et al.  (\cite{mcmahon92})  as a  pair  of  z~$=4.50$ quasars
separated by  $0.9$\arcsec.  Kochanek et al. (\cite{kochanek00}) found
that the lens  galaxy appears to  be a typical early-type lens galaxy,
with a fundamental plane   redshift estimate of   \zl$\simeq0.41$.  We
measure a significantly higher value, namely \zl$=0.632 \pm   0.002$
(Fig.~\ref{BRI0952_lens}), and the spectrum  of the lensing galaxy  is
well matched by that of an elliptical galaxy.
With this new redshift, the lensing galaxy can no longer be considered
as a member of the group of galaxies identified by Momcheva et al. (\cite{momcheva}), but
possibly of another small group in the field at z~$\simeq0.64$.

{\bf   Q~1355$-$2257  (CTQ~327)}:  This   two-image quasar,
discovered by Morgan et al. (\cite{morgan03}),  has a redshift of 
z~$=1.373$, and an image separation of $1.23\arcsec$. The redshift of the
lensing galaxies has been estimated by Morgan et al. (\cite{morgan03})
to  lie in  the  range  $0.4 <\,$\zl$\,<  0.6$. Ofek et al. (\cite{ofek})
note that their spectrum of the quasar component B shows an excess  
of emission longward of $\sim 5870$~\AA. They associate the location
of this emission excess to the 4000-\AA\ break of the lensing galaxy leading
to a redshift of \zl$=0.48$. In  a previous paper
(Eigenbrod et  al.   \cite{cosmograil3}), we  gave a   significantly  higher
tentative redshift value of \zl$=0.701$.  A re-analysis  of our data allows
to extend slightly our spectra to the blue side and to unveil the
\ion{Mg}{II} absorption line in the spectrum of the quasar 
component B. The absorption is seen at the  redshift we estimate for
the  spectrum of the lens, i.e. \zl$=0.702\pm0.001$. 
Fig.~\ref{Q1355_lens} displays the new quasar spectrum with 
an enlargement on the region around the \ion{Mg}{II} absorption.

{\bf HE~2149$-$2745}:  Wisotzki et al.  (\cite{wisotzki96}) found this
bright gravitationally  lensed quasar at  z~$=2.033$, with two images
separated by $1.71$\arcsec. They gave first  estimates of the redshift
of the lensing galaxy, refined later by Lopez et al. (\cite{lopez98}),
who  infer a probable   redshift range of $\sim0.3-0.5$.  Kochanek  et
al. (\cite{kochanek00}) give  a fundamental plane  redshift
estimate  of $0.37\le$~\zl$\le0.50$.   Burud  et  al.  (\cite{burud02})
measure the time delay between the two images ($\Delta t=103\pm12$ days) 
and report a tentative redshift of
\zl$=0.495\pm0.010$ by cross-correlating the lens spectrum 
with a template spectrum of an elliptical galaxy, but they notice that
the signal-to-noise of the  correlation function is poor.   
Our direct spectroscopic result disagrees with all previous estimates, 
with the new value of \zl$=0.603\pm 0.001$. 
This value corresponds to the photometric redshift 
z~$\simeq0.59$ (later spectroscopically confirmed at z~$=0.603$) of a galaxy group in the field 
(Williams et al. \cite{williams}, Momcheva et al. \cite{momcheva}). 
Faure et al. (\cite{faure}) also
reported this group of galaxies at the photometric redshift z~$=0.7\pm0.1$.

\section{Summary and Conclusions}

We present   straightforward VLT   spectroscopic  observations  of six
gravitationally   lensed quasars and  we  measure  the redshift of the
lensing galaxy directly from the  continuum light spectrum and several
sets of absorption lines.  The MOS  mode in which all observations are
taken  and the subsequent observation of several PSF stars are crucial
to carry out a reliable decontamination  of the lens spectrum by those
of the quasar images.  The PSF stars are also used to carry out a very
accurate flux calibration of the spectra.

Three  of our  redshifts  measurements  also correspond  to absorption
lines  seen in the  quasar spectrum. 
Following the work of Williams et al. (\cite{williams}) and
Momcheva et al. (\cite{momcheva}), we identify three lensing galaxies 
as members of small groups of galaxies. A re-analysis of  the spectra of
Q~1355$-$2257, which lens spectrum  has a low signal-to-noise ratio in
Eigenbrod   et    al.~(\cite{cosmograil3}),  allows   to  detect   the
\ion{Mg}{II} doublet in absorption  in the quasar spectrum and confirm
our previous redshift estimate, i.e. \zl$=0.702\pm0.001$.
  
Finally,  our probably most exciting results 
are our  new  redshift   measurements  for  the two lensed quasars
with time-delay determinations FBQ~0951$+$2635  and
HE~2149$-$2745, that are not compatible with previous estimates.  We find
\zl$=0.260 \pm 0.002$ instead of \zl$=0.21$ for the first and \zl$=0.603
\pm 0.001$ for the second, instead of \zl$=0.495\pm0.010$. 
This implies that previous estimates  of H$_0$ based on  these two systems  must be
revised to higher values, for a given lens model.  While the impact of
the change in redshift is   negligible for FBQ~0951$+$2635 given   the
present   uncertainties  on the   measured  time  delay  (Jakobsson et
al.~\cite{jakobsson05}),  it   is sufficiently large    in the case of
HE~2149$-$2745 (Burud  et al.~\cite{burud02}) to  justify some
new modeling of the system.

These new lens redshifts have a direct impact on several previous studies. 
More specifically  the inferred value for H$_0$ from the multiple-lens models of
Saha et al. (\cite{saha}) should be updated with these new redshifts, 
as well as with the recently measured time delay of SDSS~J1650$+$4251 
by Vuissoz et al. (\cite{vuissoz}).
In other studies (e.g. Kochanek \cite{kochanek02} \& \cite{kochanek03},
Oguri \cite{oguri}), HE~2149$-$2745 is one of lens systems that has been 
used to argue possible low H$_0$ values from time delays. 
The new lens redshift significantly increases the derived H$_0$, thereby 
weakening the possible low H$_0$ problem.

\begin{acknowledgements}
  The authors are very grateful  to the ESO  staff at Paranal for the
  particular care paid to the slit alignment necessary to perform the
  spectra deconvolutions. 
  COSMOGRAIL is financially supported by the Swiss National Science Foundation (SNSF). 
\end{acknowledgements}

\end{document}